\documentclass{elsart}

\usepackage{graphicx}
\usepackage{amssymb}

\begin{document}

\begin{frontmatter}

\title{Vector opinion dynamics in a model for social influence}

\author{M. F. Laguna}$^{1}$\thanks{Corresponding author. Fax: +54-2944-445259.}
\ead{lagunaf@cab.cnea.gov.ar},
\author{Guillermo Abramson}
\ead{abramson@cab.cnea.gov.ar}, and
\author{Dami\'an H. Zanette}
\ead{zanette@cab.cnea.gov.ar}

\address{Consejo Nacional de Investigaciones Cient\'{\i}ficas y
T\'ecnicas, Centro At\'omico Bariloche and Instituto Balseiro,
8400 Bariloche, R\'{\i}o Negro, Argentina}

\begin{abstract}
We present numerical simulations of a model of social influence, where
the opinion of each agent is represented by a binary vector. Agents
adjust their opinions as a result of random encounters, whenever the
difference between opinions is below a given threshold.  Evolution
leads to a steady state, which highly depends on the threshold and a
convergence parameter of the model. We analyze the transition between
clustered and homogeneous steady states. Results of the cases of
complete mixing and small-world networks are compared.
\end{abstract}

\end{frontmatter}

\section{Introduction}

Much attention has been paid in recent years to the potential
applications of the methods of statistical physics to complex systems
and phenomena that were traditionally considered to lie far outside
the interests of physicists \cite{mikh1}. Modeling such systems as
sets of interacting dynamical elements has proven to be a fruitful
procedure to capture the essential mechanisms that lead to generic
forms of emergent collective behavior --such as pattern formation and
synchronization \cite{mikhailov,kuramoto}-- actually observed in many
natural processes. This program has been successfully applied to a
large variety of systems in the fields of physics, chemistry, and
biology \cite{mikhailov,winfree}, and has also been extended to the
study of social and economical systems \cite{axelrod}. In
this class of systems, where elementary agents are human beings,
models aim at describing rational behavior, such as decision
making or the adoption of a strategy.

In this work we study a generic model of social influence
inspired in Axelrod's proposal \cite{axelrod}, in which the
interaction between individuals tends to homogenize the state of
the population, in the presence of some pre-existing homogeneity.
In Axelrod's model, each agent is characterized by a ``cultural''
state defined as a string of \emph{features}, each one of them
being a \emph{trait} of integer value. The agents are arranged on
a bidimensional lattice, and can interact with nearest neighbors.
The dynamical rules of the model prescribe that two agents can
interact if they have some cultural similarity, v.g. if some of
their features have equal traits. This pre-existing similarity
defines a probability that allows the active agent to further
approach the state of her partner, by adopting one of her
neighbor's cultural features. The system may evolve into an
inhomogeneous state of ``cultural regions'' whose number and
sizes depend on the parameters of the model. Interesting analysis
are given in Refs. \cite{castellano,klemm1}, where statistical
properties of the dynamics are shown.

In this context, we use the term {\it opinion} for the set of
individual attributes that are subject to social influence and
can be updated as a result of the interaction between agents.
Opinions can be modeled, in a simple approach, as binary numbers
\cite{weisbuch1} and their dynamics have been first studied
in economy \cite{folmer}. Axelrod's ``cultural'' state, restricted
to features made up of binary traits (0 or 1), corresponds to what
we call here ``opinion'' state, and has been used in Refs. 
\cite{weisbuch,deffuant}. It may be interpreted as the answers to
a survey with dichotomic (yes/no) questions aimed at defining the
agent's position on a given subject. Alternatively, opinions can
be considered as continuous variables. The conditions to reach
consensus in this case were analyzed in several works
\cite{stone,weisbuch}.

In our model, we use opinions defined by a binary string, i.e.,
the state of each agent is a vector of components 0 or 1. Since we
are interested in the propagation of opinions within a population,
agents are randomly selected in pairs, instead of the regular
lattice of Axelrod's model, which is more appropriate for the
geographical spread of culture. If the opinions of the selected
agents are close enough, they interact. The probability of
interaction is the same for all agents, and is controlled by a
threshold-like parameter. This is also in contrast with Axelrod's
model, where the probability of interaction increases strictly
with the cultural similarity. As for the underlying structure of
the population, we analyze two cases. The first corresponds to the
situation of complete mixing, where any pair of individuals can
interact and adjust their opinions. In the second, we assume that
the agents are in the nodes of a small-world network and the
interaction process occurs only between connected agents.

In the following section we introduce the model for the opinion
dynamics for the complete mixing case, and study the dynamics
of the system. In Sec. III, the properties of the stationary state are
analyzed. In Sec. IV, we introduce the small-world model and its main
results, and emphasize similarities and differences between both
cases. Finally, we summarize and discuss our most important results.

\section{Social influence and the dynamics of opinion}

Our model consist of a set of agents that update their opinions
as a result of mutual influence. We consider a population of $N$
agents, each of them characterized by a binary vector of $k$
components representing her opinion. The number of possible
different opinions is $2^k$, and $k$ is the dimension  of the opinion
space. The initial condition is a random and uniform distribution
of opinions. At each time step $t$ two randomly chosen agents
meet. If the difference between their opinions is lower than a
threshold $u$, one of them copies each opinion component from the
other with probability $\mu$. This event is called an
``interaction'' in the rest of the paper. Clearly, its occurrence
is governed by $u$, the parameter that characterizes the
existence of some --even if partial-- previous coincidence of
opinions. The probability of change of an agent's opinion is
$p=1-(1-\mu)^{d}$, where $d\leq u$ is the distance between the
opinions of the interacting pair. Thus, $\mu$ plays the role of a
convergence parameter of the model.

To quantify the difference between two opinions we use the Hamming
distance, i. e., the number of different components between the two
binary vectors. In consequence, the interaction process occurs when
agents agree in at least $k-u$ opinion components. The threshold $u$
is taken as constant in time and across the whole population.

\begin{figure}
\centering
\resizebox{\columnwidth}{!}{\includegraphics[15,100][819,582]{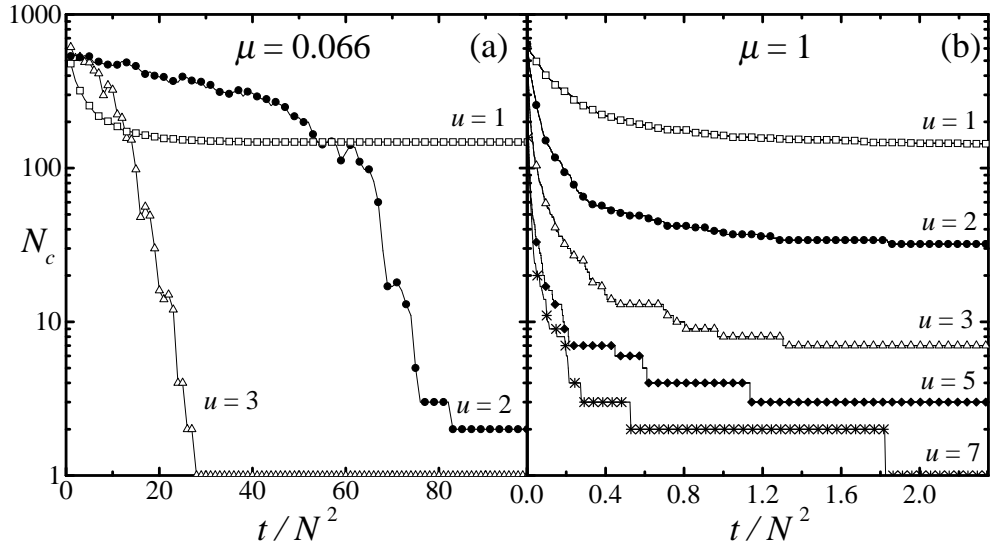}}
\caption{Number of clusters $N_c$ as a function of time for a system
of $N=1024$ agents, $k=10$ and several values of the threshold, as indicated 
in the plots. (a)  $\mu=0.066$. In this case the behavior of $N_{c}$ for $u>3$ 
is similar to the case with $u=3$. (b) $\mu=1$.}
\label{fig1}
\end{figure}

In our simulations, we consider $16 \leq N \leq 4096$ and $k=10$,
so that $2^k=1024$. Systems with $N<2^k$ are said to be
``diluted,'' whereas systems with $N>2^k$ are ``saturated,''
since there will always be at least two agents with the same
opinion. Our numerical calculation thus cover both extremes of
saturation. We have used several values of the convergence
parameter, between $\mu=0.066$ and $\mu=1$. We show here the
results corresponding to these two values, that serve well as
characterization of the two observed dynamics. As we show below,
the number of time steps needed to reach the stationary state,
$t_f$, strongly depends on $\mu$ and scales with $N^2$. We take
$t_f = 500 N^{2}$ for the lowest value $\mu$ used and $t_f = 100
N^{2}$ for the highest value.

Our analysis begins with the study of opinion evolution from a
random distribution in a well mixed population, where any two
agents are allowed to interact. We find that the system evolves,
at long times, toward clusters of homogeneous opinions. As a
characterization of the state of the system we compute, at each
time step, the number of agents with the same opinion and
calculate the number of clusters of different opinions, $N_{c}$.
In Fig. \ref{fig1} we plot the evolution of $N_c$ for several
values of the threshold. Figure \ref{fig1}(a) shows the results
for the convergence parameter $\mu=0.066$, and Fig. \ref{fig1}(b)
those for $\mu=1$. A low value of $\mu$ insures a much slower
approach to a stationary state. At short times we find that
$N_{c} \gg 1$, as expected for a system with a random
distribution of opinions. At long times, for sufficiently large
thresholds, the number of clusters is $N_{c} = 1$, i.e. full
consensus is achieved. If the threshold is low enough we have,
instead, $N_{c}>1$ at long times. This means that clusters of
different opinions exist in the stationary state, in spite of the
fact that complete mixing of the interaction mechanism and the
homogeneity of the initial configuration allow the propagation of
an opinion through the whole population. The transition from a
multi-clustered state to a single cluster one occurs at higher
values of $u$ for the fast $\mu=1$ system than for the slower
$\mu=0.066$ one. This transition is also sharp for $\mu=0.066$
(Fig. \ref{fig1}(a)) and smooth for $\mu=1$ (Fig. \ref{fig1}(b)),
a point to which we will return below.

Figure \ref{fig1}(a) shows, moreover, that the evolution in the
number of clusters is not monotonous. There are randomly
distributed events where $N_c$ increases, caused by the opinion
rearrangement that takes place before two clusters collapse. This
feature disappears when averaging over different realizations of
the system, as is the case of the rest of the results shown in
this paper.

\begin{figure}
\centering
\resizebox{\columnwidth}{!}{\includegraphics[15,100][819,582]{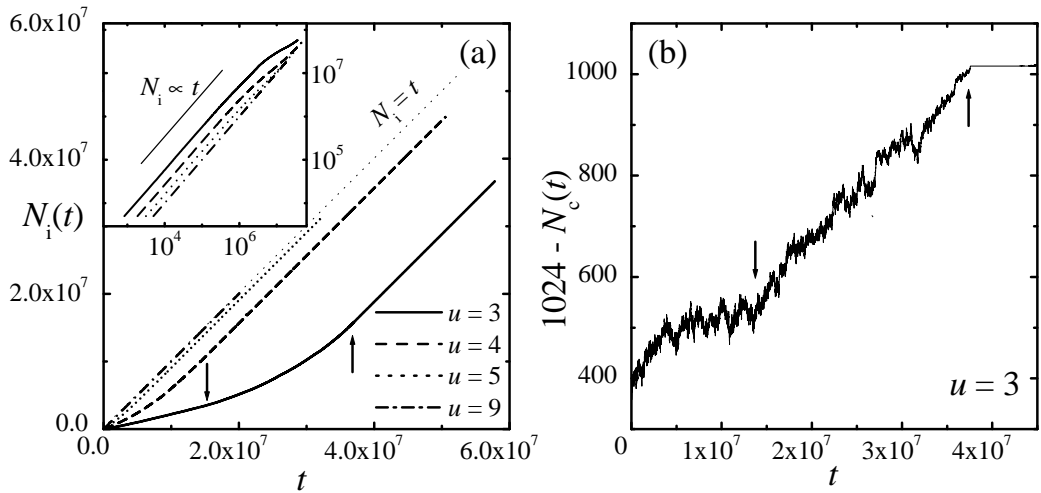}}
\caption{(a) Number of interactions $N_i$ as a function of time for a
system of $N=1024$ agents, $k=10$, $\mu=0.066$ and four values of
the threshold: $u=3$, 4, 5 and 9. Inset: The same quantities in
log-log scale. (b) Complement of the number of clusters,
$2^k-N_c(t)$, for the realization $u=3$ shown in (a). The arrows
indicate the transitions between regimes (see text). Different
realizations of the process show the same regimes, even though
their duration depend on the initial condition and the details of
the clusterization that it induces in the system.}
\label{fig2}
\end{figure}

\begin{figure}
\centering \resizebox{.8\columnwidth}{!}{\includegraphics{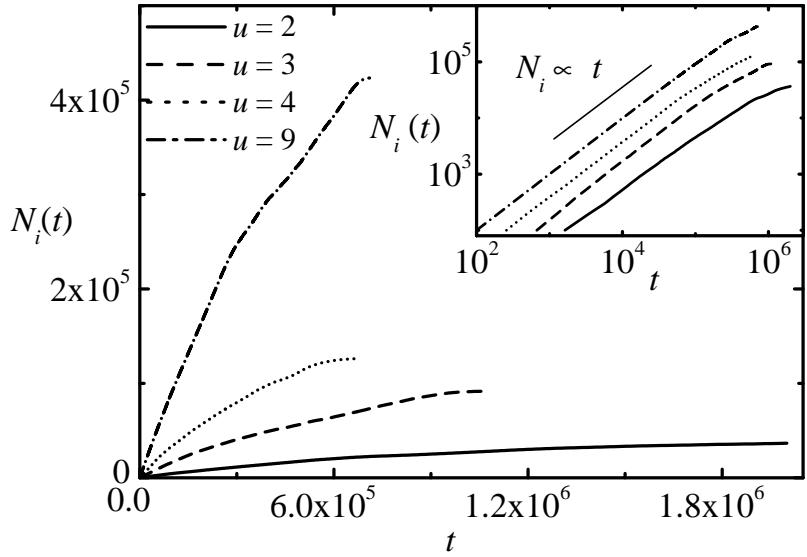}}
\caption{Number of interactions $N_i$ as a function of time for a
system of $N=1024$ agents, $k=10$, $\mu=1$ and four values of the
threshold: $u=2$, 3, 4 and 9. Inset: The same quantities in
log-log scale.}
\label{fig3}
\end{figure}

Another characterization of the opinion dynamics is provided by
the evolution of the cumulative number of interactions in the
system, $N_i(t)$. If the threshold is large (of the order of $k$),
every pair of agents selected in a given time step does interact,
i.e. $N_i(t) = t$. In the opposite limit, if $u$ is zero there
are no interactions and $N_i(t) = 0$ for all $t$. For
intermediate values of $u$, the quantity $N_i(t)$  illustrates
the dynamical state of the system. In Figs. \ref{fig2}(a) and
\ref{fig3} we plot $N_i$ as a function of time for different
values of the threshold $u$, and for $\mu=0.066$ and $\mu=1$
respectively.

In Fig. \ref{fig2}(a) we find three regimes. An initial one is
characterized by a linear growth of $N_i$, with a slope that
depends on the threshold $u$. This regime is relatively short,
and can be seen more clearly in the double logarithmic plot in
the inset of Fig. \ref{fig2}(a), where the vertical separation of
the lines represent the difference in their slopes. This
transient regime corresponds to the first stage of
clusterization. Since the initial condition is uniform over the
phase space, every agent finds her randomly picked potential
partner at a random distance, uniformly distributed over the
phase space. Due to the fact that the interaction becomes
effective only if the partner lies within a ball of radius $u$
from the agent, the probability of interactions is proportional
to the fraction of the phase space covered by such balls. If
$u\approx k$, these balls cover almost all the space and the
activity is the fastest possible, with $N_i\approx t$, as can be
seen in the figure, for the case $u=9$. For smaller $u$, the
activity is slower and the slope of $N_i(t)$ tends to 0. This
initial regime continues during the first stage of the
clusterization process, and Fig. \ref{fig2}(b) allows a
visualization of its extent. Here we show the complement of the
number of clusters, $2^k-N_c(t)$,  for $k=10$, corresponding to
the simulation run of $u=3$ shown in Fig. \ref{fig2}(a). A small
arrow indicates the end of this regime, where the clusterization
reaches a plateau. At this point, the system is composed of a
great number of clusters distributed over the phase space. Since
the convergence parameter $\mu=0.066$ allows an interacting pair
to approach their opinions without collapsing them into a single
one, most of these clusters lie close to each other. This is in
contrast to what we find for $\mu=1$, as discussed below. After
this plateau, the clusters have become near enough to interact
more often, and an acceleration of the activity appears for all
values of $u$. This is apparent in Fig. \ref{fig2}(a) as an upward
bending of $N_i(t)$, and in Fig. \ref{fig2}(b) as a nearly linear
growth of $2^k-N_c$. This mechanism rapidly reduces the number of
clusters, until there are only a few of them remaining (typically
around 10 for $u=3$). This marks the beginning of the third and
final regime, indicated by the second arrow for the case $u=3$ in
both figures. The remaining clusters are very close to each
other, and the dynamics proceeds at the maximum possible speed of
one interaction per time step until the final state of a single
opinion is reached, for all $u>2$.

In Fig. \ref{fig3} we show the curves $N_i$ vs. $t$ that
correspond to the highest convergence parameter $\mu=1$. For all
values of $u$ we find two regimes, a linear one at short times
with a slope dependent of $u$ (for the same underlying reasons as
in the case $\mu=0.066$), and a long time behavior with a slower
increase with time. After the initial linear regime in which
there is an intense activity, the interactions become more spaced
in time and finally stop, when the stationary state is reached.
The reason for this slowing down, at variance with the
acceleration displayed when $\mu=0.066$, resides in the extreme
value of the convergence parameter. Since $\mu=1$, the opinions
of a pair of interacting agents become identical after the
interaction. The clusters, in consequence, become effectively
isolated from one another, and intermediate opinions do not exist
after the first clusterization stage . Eventually, and depending
on the value of $u$, a final state with several clusters becomes
stationary.

\begin{figure}
\centering \resizebox{.8\columnwidth}{!}{\includegraphics{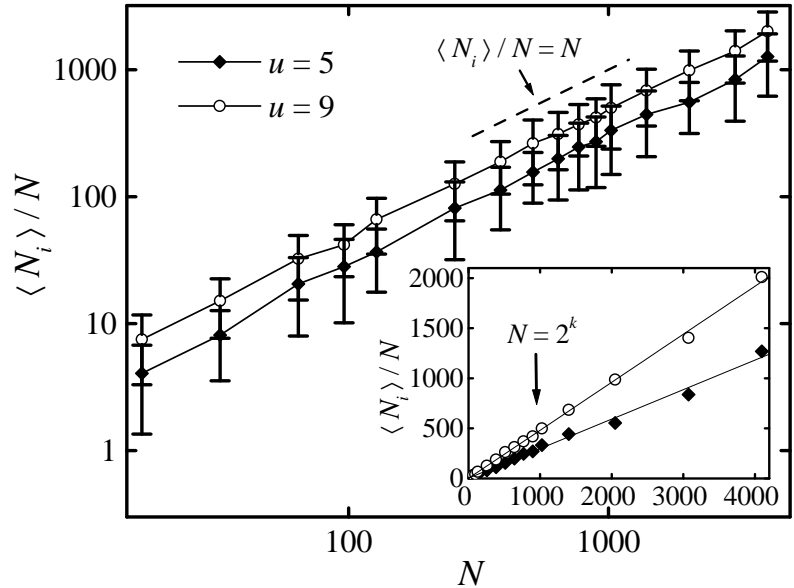}}
\caption{Total number of interactions as a function of system
size, averaged over 100 realizations, with $k=10$, $\mu=1$ and two
values of the threshold: $u=5$ (diamonds), $u=9$ (open circles).
Inset: linear-linear  plot of the same data; lines are linear fits.}
\label{fig4}
\end{figure}

In Fig. \ref{fig4} we plot the average over $100$ realizations of
the total number of interactions $N_{i}$ needed to achieve the
stationary state, as a function of the system size $N$ for two
values of the threshold $u$, $k=10$ and $\mu=1$. We observe an
approximately linear increase of $\langle N_{i}\rangle$ with
$N^2$, with a slope depending on the threshold, as shown in the
inset. For all system sizes we find the same dependence with $N$,
which coincides with the one observed for the time $t_f$ needed to
reach the stationary state. This means that the ratio $N_i/t_f$,
measuring the average number of interactions per time unit during
the complete evolution of the system, is independent of the
system size. For $\mu=0.066$ the dependence on $N$ is the same,
even though there is a factor between 10 and 100 in the total
number of interactions needed to reach the final state. Since the
behavior of $\langle N_{i}\rangle$ does not depend on $N$ in a 
qualitative way, we use an intermediate representative value, 
$N=2^k$, in the rest of the paper.

\section{Characterization of stationary states}

As shown in the previous section, a stationary state is achieved
after a number of interactions that depends on the convergence
parameter $\mu$ and on the particular realization. We are
interested now in the dependence of the final state of the system
on the threshold and the convergence parameter.

\begin{figure}
\centering \resizebox{.8\columnwidth}{!}{\includegraphics{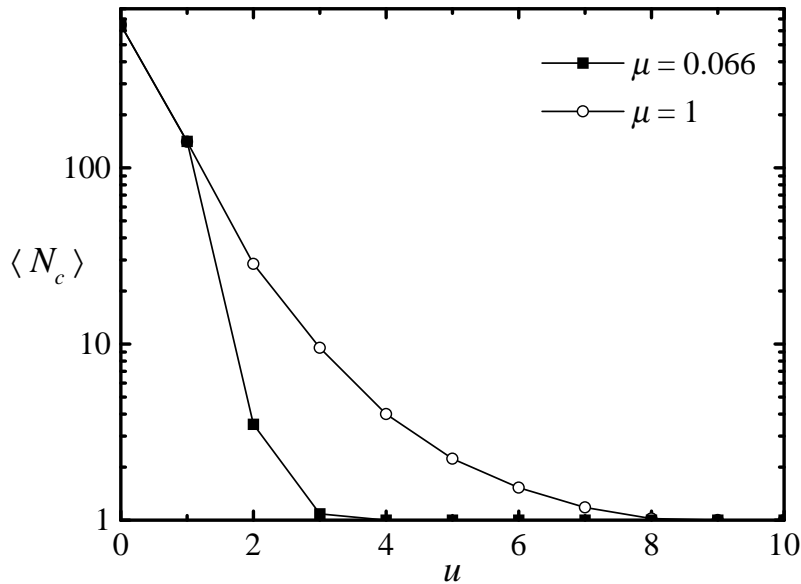}}
\caption{Average number of clusters $\left<N_{c}\right>$
as a function of the threshold, for 100 realizations with $N=1024$
and two values of the convergence parameter: $\mu=0.066$
(squares) and $\mu=1$ (open circles).}
\label{fig5}
\end{figure}

\begin{figure}
\resizebox{\columnwidth}{!}{\includegraphics[15,100][819,582]{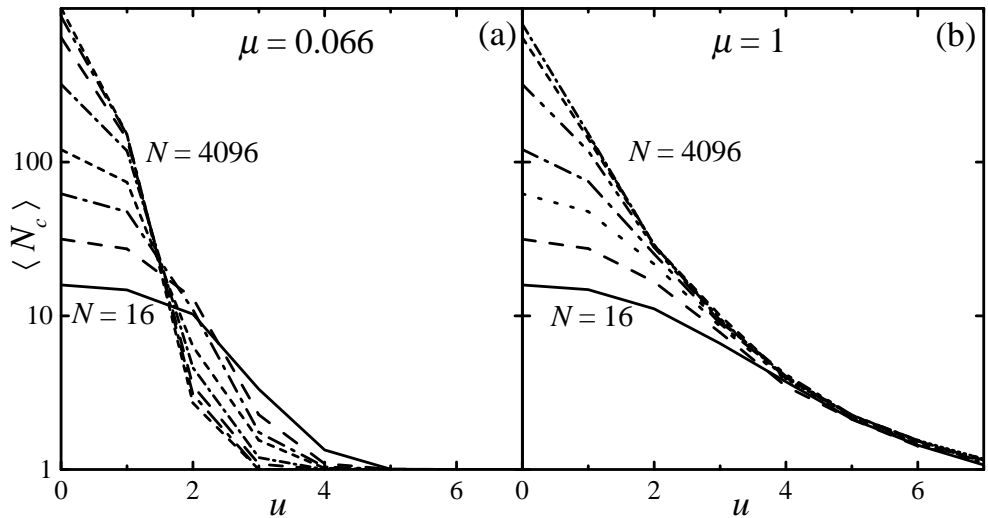}}
\caption{Average number of clusters $\left<N_{c}\right>$
as a function of the threshold for 200 realizations and two values
of $\mu$, as indicated in the plots. Different curves correspond
to system sizes $N=16$, $32$, $64$, $128$, $384$, $1024$, $2048$, and
$4096$.}
\label{fig6}
\end{figure}

In Fig. \ref{fig5} we plot the number of clusters in the
stationary state, $\left<N_{c}\right>$, as a function of the
threshold $u$ for two values of the convergence parameter,
$\mu=0.066$  and $\mu=1$. Each dot of the curve corresponds to an
average over $100$ realizations. In both cases the value of
$\left<N_{c}\right>$ for $u<2$ is very similar and corresponds to
a situation where several clusters of different opinions coexist
in equilibrium. In the region $u>7$, both cases are also similar,
with $\left<N_c\right>=1$. This is a situation where a single
opinion always prevails. In the region $2<u<7$ the mean number of
clusters depends on the convergence parameter. The transition
from the multi-cluster state to the single opinion state is
smooth for $\mu = 1$, whereas for $\mu=0.066$ there is a sharp
transition at $u\lesssim 3$. For the system size shown here this
last transition is found for $\mu < 0.99$. A signature of the
deep difference between the two transitions is their behavior in
a finite size analysis. We show this in Fig. \ref{fig6} for
$\mu=0.066$ and $\mu=1$, for a range of system sizes that
correspond to increasing saturation, since $k$ remains fixed. It
is clear that the transition becomes sharper with increasing
system size when $\mu<1$, and that it preserves its smoothness
when $\mu=1$. This is an indication that there is a further
transition at $\mu=1$ between the two types of behavior. Only the
fastest systems ($\mu=1$) allow the persistence of multi-opinion
populations at intermediate values of the threshold $u$. In Figs.
\ref{fig6}(a) and (b) we show a diversity of behaviors that
correspond to the different regimes of saturation. This analysis
does not allow a further characterization of the transitions.
Indeed, the thermodynamic limit would require not only that
$N\to\infty$ but also that $k\to\infty$, at a fixed saturation in
opinion space. Such systems, with opinions defined by infinite
features, go beyond the interest of the present study. In a
related context, this has been carried out in Ref.
\cite{castellano}, where the switch from the monocultural to the
multicultural states in Axelrod's model is identified as a phase
transition. The roles of noise and of network structure have been
analyzed in Refs. \cite{klemm1,klemm2}.

\begin{figure}
\centering \resizebox{.8\columnwidth}{!}{\includegraphics{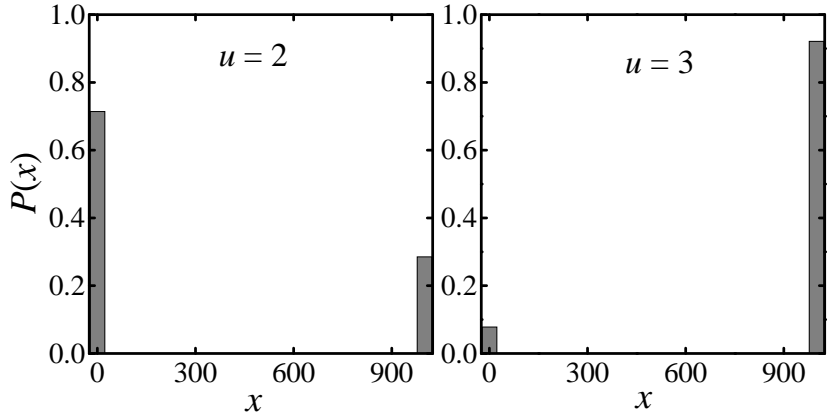}}
\caption{Distribution of cluster population $P(x)$,
for 200 realizations of a system with $N=1024$, $\mu=0.066$ and
two values of the threshold ($u=2$ and $3$).}
\label{fig7}
\end{figure}

The transition as a function of the threshold can be further
characterized analyzing the distribution of the population of
clusters, $P(x)$, for different values of $u$. If the system is in
a homogeneous state, all the agents having the same opinion, then
$N_{c}=1$ and the distribution has a peak at $x \simeq N$. A
situation of many clusters with a few agents in each cluster is
characterized by a distribution $P(x)$ with a dominant peak in $x
\sim 1$. In Fig. \ref{fig7} we show the results for the case
$\mu=0.066$ and two values of $u$ near the transition. For $u<2$
the histogram has a single peak in $x=1$ whereas for $u>3$ it has
a single peak in $x=N$. The threshold at which the transition
between a clusterized state and a homogeneous state takes place is
between $u=2$ and $3$. This result coincides with that of Fig.
\ref{fig5}.

\begin{figure}
\centering \resizebox{.8\columnwidth}{!}{\includegraphics{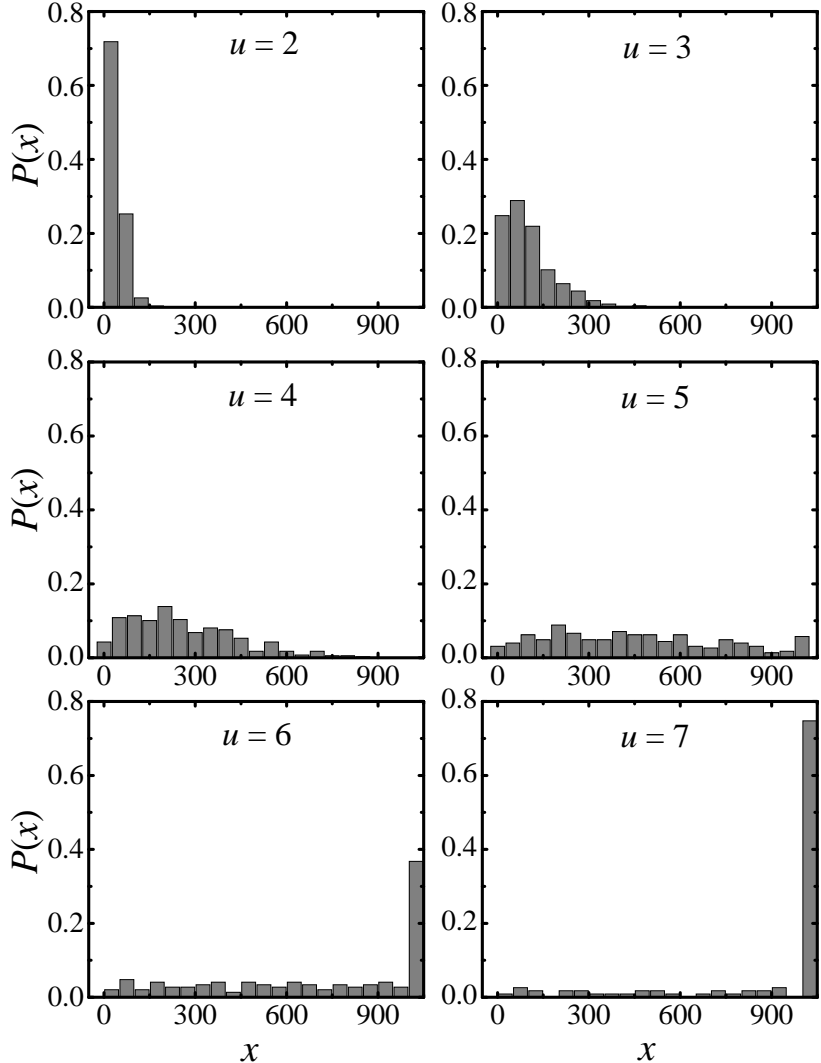}}
\caption{Distribution of cluster population $P(x)$, for 100
realizations
of a system with $N=1024$, $\mu=1$ and six values of the threshold, as
indicated in each plot.}
\label{fig8}
\end{figure}

A very different sequence of histograms is obtained for $\mu=1$,
as we show in Fig. \ref{fig8}. The transition from a highly
clustered state (the histogram with a peak in $x=1$, at $u=2$) to
a homogeneous state (the histogram with a peak in $x=N$, at $u=7$),
takes place gradually for intermediate values of $u$. This result
is also in agreement with the behavior observed in the $\left<
N_c \right>$ curve.

Note that to describe completely the state of the system we need
both quantities, $\left< N_c \right>$ and the sequence of
histograms. As an example, observe that in Fig. \ref{fig5},
$\left< N_{c} \right> \sim 4$ for both $u=2$ with $\mu=0.066$ and
$u=4$ with $\mu=1$. However, the histograms are very different in
Fig. \ref{fig8} with $u=2$ and Fig. \ref{fig9} with $u=4$.

The results of this section indicate that a well-defined
transition between a homogeneous state and a highly clustered
state takes place in this model as a function of $u$. A
complementary result is that the parameter $\mu$ not only
modifies the convergence speed towards equilibrium but also
changes the structure of the stationary state. This last result
contrasts with Ref. \cite{weisbuch1}, where the authors report
that the convergence parameter only modifies the time needed to
reach the equilibrium.

\section{Complex mixing vs. small-world network}

The two previous sections have dealt with the case of complete mixing,
in which any agent can interact with any other one in the system. How do
the dynamical and steady states change if the underlying structure of
the population is different? Here we analyze the case in which the
agents are situated at the nodes of a small-world network
\cite{watts}, which represents more accurately some aspects of real
social populations. At variance with a complete mixing system, a
small-world network exhibits some degree of local clustering in the
neighborhood of each agent. We introduce the parameter $p$, the
reconnection probability, that measures the randomness of the
small-world and interpolates between an ordered lattice ($p=0$) and a
random graph ($p=1$). Another property that characterizes the
small-world is the network connectivity $K$. The small-world network
is built from a one-dimensional regular lattice in which each node is
linked to its $2 K$ nearest neighbors. Then, each link is rewired
with probability $p$ to a randomly chosen node. Multiple links are
forbidden and disconnected networks are discarded. Every agent in the
small-world network has, on the average, $2 K$ nearest
neighbors.

The interaction process occurs only between linked sites. Now, we
randomly choose an agent and then we select, also at random, one of
his neighbors to interact with. The mechanism of the interaction is
the same as in the complete mixing model. To ease the comparison with
previous results, we use $N=1024$ and $k=10$ in all the figures of
this section.

\begin{figure}
\centering \resizebox{.8\columnwidth}{!}{\includegraphics{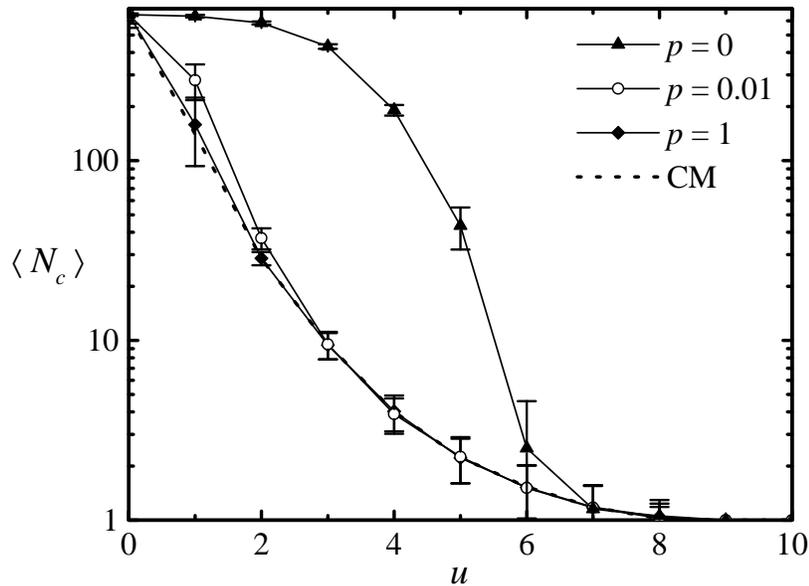}}
\caption{Average number of clusters $\left< N_c \right>$ as a function
of the threshold for 500 realizations with $N=1024$, $\mu=1$ and
different values of the reconnection probability $p$ in a
small-world network with $K=3$. We compare this results
with the complete mixing (CM) case.}
\label{fig9}
\end{figure}

Figure \ref{fig9} shows the average number of clusters $\left< N_c
\right>$ as a function of the threshold $u$ for a small-world network
with $K=3$, $\mu=1$ and three different values of the reconnection
probability $p$. We compare this results with those
obtained in the complete mixing case of Fig. \ref{fig5} (open
circles). The behavior of $\left< N_c \right>$ for $p \neq
0$ is very similar to the results of the complete mixing case in all
the range of thresholds. The only qualitatively different situation
is the case $p=0$, that corresponds to the agents in a regular
lattice. In this case, and for $u<7$, the number of clusters with
different opinions is significantly larger than those corresponding
to other values of $p$. The regular lattice --and only the one which
is completely regular-- is less effective  at disseminating the
opinions. The same dependence with $p$ was observed for the case of
$\mu=0.066$ (not shown here).

\begin{figure}
\centering \resizebox{.8\columnwidth}{!}{\includegraphics{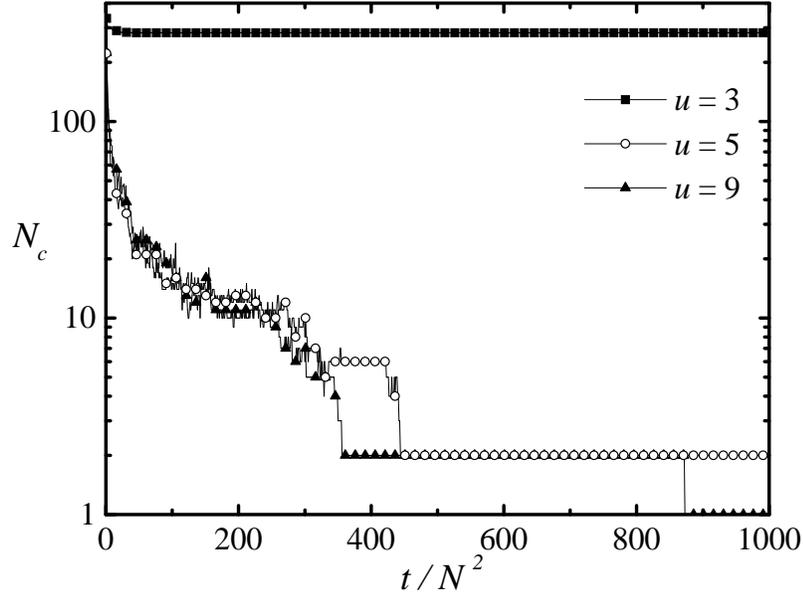}}
\caption{Number of clusters $N_c$ as a function of time, for a
system arranged on a regular lattice of  $N=1024$ agents, $k=10$,
$\mu=0.066$ and three values of the threshold. This situation
corresponds to a small-world network with $p=0$.}
\label{fig10}
\end{figure}

\begin{figure}
\resizebox{\columnwidth}{!}{\includegraphics[15,100][819,582]{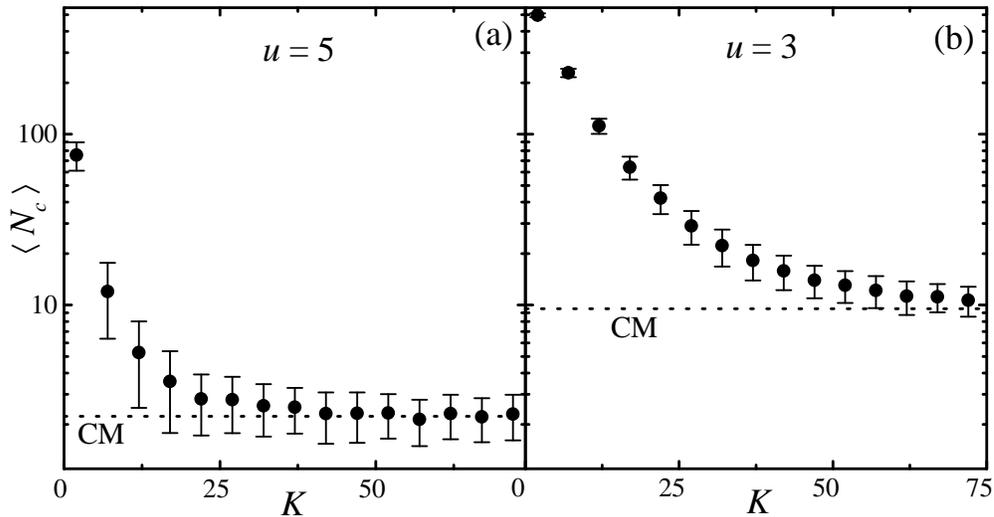}}
\caption{Mean number of clusters as a function of network connectivity
$K$ for 200 realizations in a small-world with $p=0$, $\mu=1$ and 
$N=1024$. (a) The threshold is $u=5$  (b) In this picture, $u=3$. 
Dashed horizontal lines indicate the complete mixing (CM) case.}
\label{fig11}
\end{figure}

These results indicate that the dynamics in a small-world
configuration does not differ from that of a well mixed system,
and the only different case is the one corresponding to $p=0$,
actually, not a small-world network but a regular lattice. The
transient evolution of single realizations of such a system is shown
in Fig. \ref{fig10}, where we plot $N_c(t)$ for three values of the
threshold. The parameters are the same as in Fig. \ref{fig1}, where
we use the lowest value of $\mu$ to have the time evolution in
detail. Note that the system on the lattice needs roughly $10$
times more iterations than the completely mixed one. We also
observe the random fluctuations originated by opinion
rearrangements in the clustering process.

Finally, in order to analyze the effect of the connectivity in
the system, we change the values of $K$. As can be observed
in Fig. \ref{fig11}, the mean number of clusters tends to the
complete mixing case at high values of $K$. The number of
nearest neighbors necessary to recover that behavior depends
on the threshold.

\section{Conclusion}

We have studied a generic model of social influence in which
agents adjust their opinions as a result of random binary
encounters whenever their difference in opinion is below a given
threshold.

We have analyzed the dynamical behavior of the system and found that 
the cases $\mu=1$ and $\mu < 1$ behave differently.
For the extreme case $\mu=1$ we observe an initial linear regime 
of intense activity that ceases when intermediate opinions disappear. 
This first stage is followed by a different regime in which the 
interactions become more spaced in time because the opinion clusters 
are effectively isolated, even for $u$ very near $k$.
Finally, a saturation occurs when the stationary state is reached.
The slower case $\mu=0.066$ shows three regimes. 
The first stage of clusterization is similar to the
one observed in the previous case, but it produces a great number of clusters that lie close to each other.
A second stage starts, in which the clusters interact more often, giving rise to an acceleration of the activity until a few clusters remain in the system. Finally, in the third stage the dynamics proceeds at the maximum speed due to the interaction between clusters that are close enough. The dynamical process stops when a single opinion is reached (for $u > 2$) or several isolated clusters are formed (for $u < 2$).
We can see, then, that the speed of the convergent dynamics plays a relevant role not only in the final composition of the population but also in the intermediate stages of evolution.

We also found that both the total number of interactions
between individuals and total time needed to reach the
equilibrium, increase quadratically with the size of the system.
Consequently, the ratio between these two quantities,
$N_{i}/t_{f}$, is independent of the system size. 
We obtained this result for all the values of $\mu$ used.

As for the stationary state properties, we have found that the
interaction between individuals restricted by a proximity
threshold results into clustering of opinions with a number of
clusters decreasing with the increase of the threshold. Our model
presents a transition driven by the threshold, from a state with
homogeneous opinion (consensus) to a phase in which several
clusters with different opinions coexist. The nature of the
transition depends on the convergence parameter $\mu$, that changes the
structure of the stationary state. This was also observed in the
sequence of histograms of population of clusters near the
transition. Finite size scaling supports the existence of such
transition.

Finally, we have studied the influence of the underlying structure
on the behavior of the system. We found that small-world networks
and the complete mixing case have a very similar behavior, except
for the limiting case of regular graphs ($p=0$ in the small-world
model). This last case is less effective to disseminate the
opinions than the others. The peculiarity of the case $p=0$ is
found also in other contexts, within the study of small-world
networks. There has been found that the small-world phenomenon
appears at any value of the network disorder $p$, for
sufficiently large systems. Geometrical properties such as the
average path length \cite{barrat,newman}, and some dynamical
ones, such as the ferromagnetic properties of an Ising model
\cite{barrat}, have been shown to behave critically at $p=0$.
This is not the case in some other dynamical models, such as the
propagation of an epidemic \cite{kuperman}, and of a rumor
\cite{zanette}, that behave critically at a finite value of $p$
in the thermodynamical limit. In our model of opinion dynamics,
the effect of disorder is even stronger, since any nonzero value
of $p$ induces a behavior that is indistinguishable from both the
fully disordered $p=1$ case and the complete mixing case. We also
found that greater network connectivity gives rise to a decrease
in the number of clusters. The number of connected neighbors
needed to reach the complete mixing ones depends on the threshold.

Several open problems associated with opinion dynamics are worth
mentioning. For instance, the interplay between long-range effects,
such as mass-media broadcasting of individual or collective opinions,
and local interactions is an essential mechanism in modern opinion
formation. Boundary effects and inhomogeneity conditions, including
spatial variation of population density, may also strongly affect
the dynamics of social influence and the properties of the resulting
stationary state. Finally, the effects of noise, communication errors,
and in general random fluctuations, should not be disregarded in any
realistic representation of information spreading. The present model
is expected provide a versatile tool to analyze these questions.

\section*{Acknowledgement}

M.F.L. thanks the Solid State Group of Centro At\'omico Bariloche
for the computational facilities.

\end{document}